\documentclass[conference]{IEEEtran}

\IEEEoverridecommandlockouts

\usepackage[utf8]{inputenc}
\usepackage[T1]{fontenc}
\usepackage{cite}
\usepackage{inconsolata}
\usepackage{amsmath}
\usepackage{amssymb}
\usepackage{booktabs}
\usepackage{array}
\usepackage{xcolor}
\usepackage{listings}
\usepackage{float}
\usepackage{tikz}
\usetikzlibrary{positioning,arrows.meta}
\usepackage[hidelinks]{hyperref}
\usepackage{url}

\usepackage{comment}

\definecolor{rocqkw}{RGB}{0,0,0}
\definecolor{rocqcm}{RGB}{96,103,112}

\lstdefinestyle{rocq}{%
  basicstyle=\ttfamily\footnotesize,
  keywordstyle=\color{rocqkw}\bfseries,
  commentstyle=\color{rocqcm}\itshape,
  morekeywords={Definition,Theorem,Lemma,Fixpoint,Record,Inductive,Proof,Qed,
    forall,exists,match,with,end,if,then,else,fun,let,in,Prop,Type,nat,list,
    option,Some,None,Permutation,NoDup},
  columns=fullflexible,
  keepspaces=true,
  showstringspaces=false,
  breaklines=true,
  frame=single,
  framerule=0.3pt,
  rulecolor=\color{black!35},
  framesep=5pt,
  xleftmargin=4pt,
  xrightmargin=4pt,
  aboveskip=4pt,
  belowskip=2pt,
}
\newcommand{\yes}{$\checkmark$}

\newcommand{\na}{n.a.}

\newcolumntype{L}[1]{>{\raggedright\arraybackslash}p{#1}}

\makeatletter
\newcommand{\linebreakand}{%
  \end{@IEEEauthorhalign}
  \hfill\mbox{}\par
  \mbox{}\hfill
  \begin{@IEEEauthorhalign}
}
\makeatother

\begin{document}
\bstctlcite{BSTcontrol}

\title{A Comparative Formal Verification Framework for General-Purpose CPU Scheduling in Rocq}


\author{
\IEEEauthorblockN{Arooj Fatima}
\IEEEauthorblockA{
\textit{School of Interdisciplinary Engineering and Sciences}\\
\textit{National University of Sciences and Technology}\\
Islamabad, Pakistan\\
afatima.bsbi23sines@student.nust.edu.pk
}
\and
\IEEEauthorblockN{Saad Saghir Minhas}
\IEEEauthorblockA{
\textit{School of Electrical Engineering and Computer Science}\\
\textit{National University of Sciences and Technology}\\
Islamabad, Pakistan\\
sminhas.bscs23seecs@seecs.nust.edu.pk
}
\linebreakand
\IEEEauthorblockN{Adnan Rashid}
\IEEEauthorblockA{
\textit{School of Electrical Engineering and Computer Science}\\
\textit{National University of Sciences and Technology}\\
Islamabad, Pakistan\\
adnan.rashid@seecs.nust.edu.pk
}
}

\IEEEpubid{%
\makebox[\columnwidth]{%
\footnotesize
979-8-3195-0250-6/26/\$31.00~\copyright~2026 IEEE
\hfill}%
\hspace{\columnsep}%
\makebox[\columnwidth]{}%
}

\maketitle

\begin{abstract}
CPU scheduling determines which runnable process receives the processor and directly affects system responsiveness, fairness, and overall performance. Although many scheduling policies have been extensively studied and implemented, their correctness is typically justified through informal reasoning, examples, or simulations rather than machine-checked proofs. Existing formal verification efforts also focus predominantly on real-time scheduling, leaving general-purpose scheduling policies comparatively underexplored. This paper formalizes six general-purpose scheduler policies in Rocq: FIFO, SJF, Lottery, MLFQ, Stride, and CFS. By policy, we mean the rule used to choose the next runnable process and, for state-transforming models, update scheduler state. Beyond formalizing the schedulers individually, we compare their verified properties and identify recurring proof patterns across scheduler families. FIFO, SJF, Stride, and list-based CFS share a minimum-key proof pattern, MLFQ follows a layered-priority proof pattern, and Lottery follows a ticket-interval proof pattern. We formalize these proof patterns in Rocq, enabling proof reuse while allowing each scheduler to retain its own mathematical model. To the best of our knowledge, this work provides the first interactive-theorem-prover formalizations of MLFQ, Lottery, and Stride schedulers, together with CFS-style virtual-runtime scheduling as standalone general-purpose scheduler policies. We also include a CFS variant based on Rocq's verified red-black-tree library. The entire development is checked by Rocq without unfinished proofs or additional trusted assumptions, i.e., every theorem is accepted by the Rocq kernel.
\end{abstract}


\begin{IEEEkeywords}
formal verification, interactive theorem proving, Rocq, CPU scheduling, reusable proof patterns, formal modeling
\end{IEEEkeywords}


\section{Introduction} \label{SEC:Intro}
CPU scheduling is a fundamental operating system function that determines which runnable process is selected for execution on the processor. Because scheduling decisions are made continuously during system execution, they directly influence system responsiveness, throughput, fairness, processor utilization, and the overall efficiency of computing platforms. Consequently, CPU scheduling remains a core component of modern operating systems, including desktop systems, servers, cloud infrastructures, and other general-purpose computing environments.

Over the past several decades, numerous scheduling policies have been proposed to achieve different performance objectives, such as minimizing response time, improving fairness, or providing proportional resource allocation. Although these policies have been extensively studied and widely implemented, their correctness is typically justified through informal reasoning, illustrative examples, empirical testing, or simulation-based evaluation. While these approaches provide valuable insights into scheduler behavior, they cannot establish that a scheduling policy satisfies its intended correctness properties for every valid execution state.

Formal verification addresses this limitation by constructing mathematical models and machine-checked proofs that establish whether a system satisfies a precise specification for all valid states~\cite{formal-verification}. Among the available formal verification techniques, interactive theorem proving provides a particularly high level of assurance by requiring every proof step to be verified by a trusted logical kernel. In this work, we employ Rocq (formerly Coq), a mature interactive theorem prover based on dependent type theory~\cite{rocq}. Scheduler models, correctness properties, and their proofs are expressed within Rocq's logic, and every theorem is accepted only after successful kernel verification, eliminating incomplete proofs and unchecked reasoning.

Existing machine-checked verification efforts on CPU scheduling have focused predominantly on two directions. The first concerns real-time schedulability analysis, where the primary objective is to establish timing guarantees for real-time task sets~\cite{prosa}. The second verifies scheduling mechanisms as components of complete operating-system kernels~\cite{sel4,certikos}. In contrast, comparatively little attention has been devoted to machine-checked formalizations of general-purpose CPU scheduling policies as standalone mathematical models. Such models facilitate rigorous reasoning about scheduling algorithms themselves, independently of operating-system implementation details, while also enabling systematic comparison of different scheduling strategies within a common verification framework.

This paper presents a comparative formal verification framework for six representative general-purpose CPU scheduling policies in Rocq, namely First-In-First-Out (FIFO), Shortest Job First (SJF), Lottery, Multi-Level Feedback Queue (MLFQ), Stride, and Completely Fair Scheduler (CFS). For CFS, we additionally formalize a variant whose runqueue is implemented using Rocq's verified red-black-tree library. Beyond verifying each scheduler individually, we systematically compare their verified properties and identify recurring verification patterns that arise across different scheduler families. Specifically, FIFO, SJF, Stride, and list-based CFS exhibit a common minimum-key pattern, MLFQ follows a layered-priority pattern, and Lottery follows a ticket-interval pattern. We formalize these recurring verification patterns in Rocq, enabling proof reuse while allowing each scheduler to retain its own mathematical model.

The main contributions of this paper are as follows:

\begin{itemize}
\item We present machine-checked Rocq formalizations of six representative general-purpose CPU scheduling policies: FIFO, SJF, Lottery, MLFQ, Stride, and CFS. For CFS, we additionally formalize a variant based on Rocq's verified red-black-tree library.

\item We establish machine-checked correctness results for the formalized schedulers, including selection safety, determinism, work conservation, starvation-freedom variants, and fairness-related properties where supported by the scheduling model.

\item We identify recurring verification patterns across scheduler families, showing that FIFO, SJF, Stride, and CFS share a minimum-key pattern, MLFQ follows a layered-priority pattern, and Lottery follows a ticket-interval pattern.

\item We formalize these verification patterns in Rocq, enabling systematic comparison and proof reuse across diverse scheduling policies without imposing a common implementation model.
\end{itemize}

\begin{figure}[!ht]
\centering
\begin{tikzpicture}[
  node distance=4.5mm,
  every node/.style={font=\footnotesize},
  box/.style={
    draw,
    rounded corners=0pt,
    align=center,
    minimum width=7.0cm,
    minimum height=7mm
  },
  arr/.style={-Latex}
]
\node[box] (models) {
  \textbf{Scheduler Models}\\
  Process, state, one-step rule\\
  FIFO, SJF, Lottery, MLFQ, Stride, CFS, CFS (RB-tree)
};

\node[box, below=of models] (props) {
  \textbf{Verified Properties}\\
  Selection safety, determinism,\\
  work conservation, fairness, starvation freedom
};

\node[box, below=of props] (patterns) {
  \textbf{Shared Proof Patterns}\\
  Minimum-key, layered-priority, ticket-interval
};

\node[box, below=of patterns] (matrix) {
  \textbf{Property Matrices}\\
  Proved, partial, open, or not applicable
};

\draw[arr] (models) -- (props);
\draw[arr] (props) -- (patterns);
\draw[arr] (patterns) -- (matrix);
\draw[arr] (props.west) to[out=180,in=180,looseness=0.8] (matrix.west);

\end{tikzpicture}
\caption{Overview of the Verification Workflow}
\label{fig:pipeline}
\end{figure}
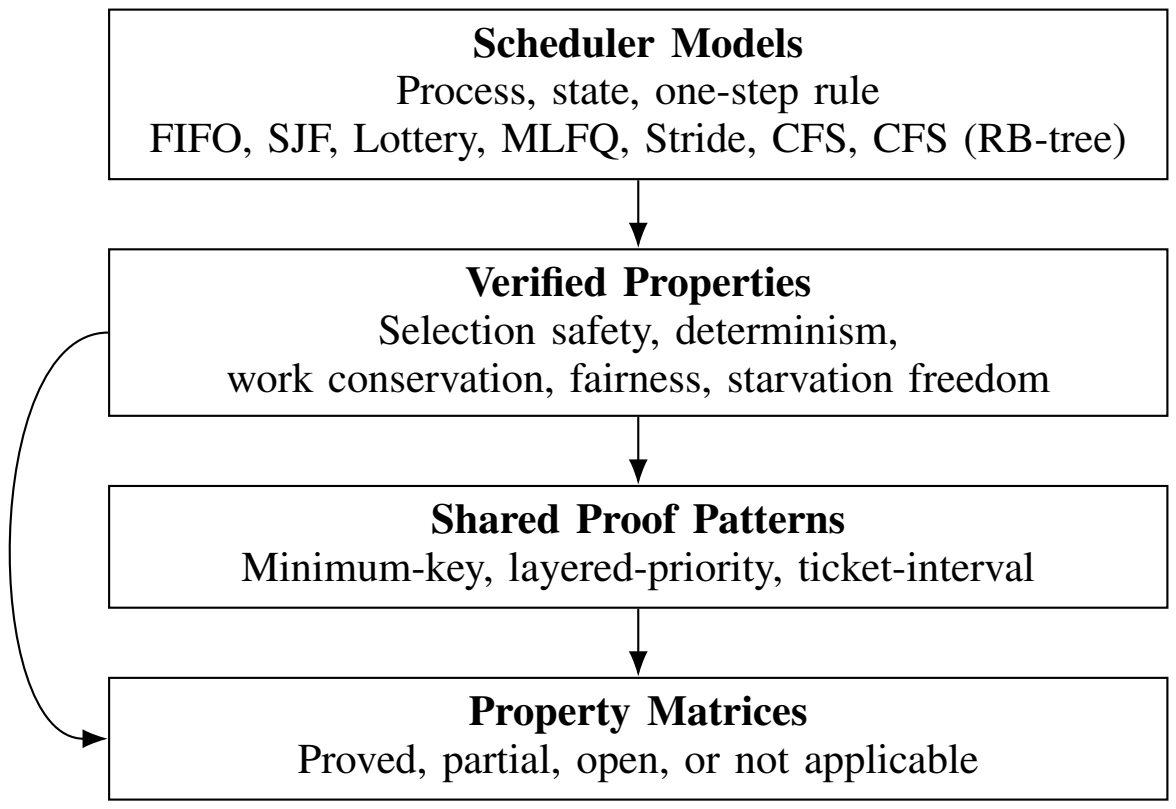

Figure~\ref{fig:pipeline} illustrates the overall verification workflow. Beginning with formal scheduler models, the framework establishes scheduler-specific correctness properties before extracting reusable verification patterns that support systematic comparison across different scheduler families. The resulting property matrices summarize which correctness properties are fully verified, partially established, left as future work, or are not applicable to a particular scheduling policy.

The remainder of this paper is organized as follows. Section~\ref{SEC:back_rel_work} reviews the necessary background and related work. Section~\ref{SEC:Modeling_frame} presents the proposed modeling framework, while Section~\ref{SEC:Schedulers} describes the formalization of the scheduling policies. Section~\ref{SEC:properties} discusses the verified properties and comparative property matrices, and Section~\ref{SEC:Interfaces} introduces the shared proof patterns. Finally, Sections~\ref{SEC:limits} and~\ref{SEC:conclusions} discuss the limitations of the current framework and conclude the paper.


\section{Background and Related Work} \label{SEC:back_rel_work}

This section provides the background necessary for the remainder of the paper. We first introduce the scheduler families considered in this work and motivate their selection. We then review the closest machine-checked verification efforts on scheduling and position our contribution with respect to the existing literature.

\subsection{Scheduler Policies}

The schedulers considered in this paper represent four fundamental approaches to CPU scheduling: ordering by a scheduling key, dynamically adjusting priorities through execution feedback, allocating processor time using proportional-share mechanisms, and approximating fair processor allocation through runtime accounting.

FIFO and SJF are representative ordering-based schedulers. FIFO selects processes according to their arrival order, whereas SJF prioritizes the process with the shortest estimated CPU burst~\cite{ostep-sched}. MLFQ represents feedback-driven scheduling by dynamically adjusting process priorities according to their observed execution behavior, thereby adapting to changing workloads without requiring prior knowledge of job lengths~\cite{ostep-mlfq}. Lottery and Stride constitute the two classical proportional-share scheduling policies. Lottery scheduling allocates processor time using randomized ticket selection, while Stride scheduling provides a deterministic proportional-share alternative based on pass values~\cite{lottery,stride,ostep-lottery}. Finally, CFS approximates fair processor allocation through virtual-runtime accounting and has served as the default scheduler in the Linux kernel for many years~\cite{cfsdoc}.

These schedulers were selected because they collectively capture the major scheduling paradigms employed in modern general-purpose operating systems while exhibiting substantially different verification structures. In particular, FIFO, SJF, Stride, and CFS naturally follow a minimum-key selection principle, whereas MLFQ and Lottery require fundamentally different reasoning based on layered priorities and ticket intervals, respectively. This diversity makes them well suited for evaluating the proposed comparative verification framework.

\subsection{Formal Verification of Scheduling}

Machine-checked verification of scheduling algorithms has primarily focused on two research directions: real-time schedulability analysis and verified operating-system kernels.

The first direction concerns real-time scheduling, where the objective is to establish timing guarantees for time-critical task sets. Prosa provides machine-checked schedulability analyses in Rocq for scheduling policies such as Earliest Deadline First (EDF) and fixed-priority scheduling~\cite{prosa}. Similarly, a Coq-based case study verifies response-time analysis for FIFO scheduling in a sporadic real-time setting~\cite{fiforta}. These efforts establish timing properties rather than the functional correctness of general-purpose scheduling policies.

The second direction focuses on verified operating-system kernels. The seL4 project proves the functional correctness of a microkernel, including its scheduler~\cite{sel4}. CertiKOS verifies certified kernel abstractions~\cite{certikos}, while RT-CertiKOS extends verified kernel development with formally verified real-time scheduling support~\cite{rtcertikos}. In these systems, scheduling forms one component of a much larger kernel verification effort rather than the primary object of verification.

We also identified an educational TLA+ specification of MLFQ derived from OSTEP~\cite{mlfq-tla}. Although it provides a useful formal specification for understanding scheduler behavior, it does not establish machine-checked correctness proofs within an interactive theorem prover.

Table~\ref{tab:related} summarizes representative formal verification efforts on CPU scheduling. Unlike previous work, which primarily targets real-time schedulability or kernel correctness, this paper focuses on machine-checked verification of general-purpose scheduling policies as standalone mathematical models while identifying reusable verification patterns that enable systematic comparison across scheduler families.

\vspace*{-0.2cm}

\begin{table}[H]
\caption{Comparison of Representative Formal Scheduling Verification}
\label{tab:related}
\centering
\footnotesize
\begin{tabular}{@{}L{1.85cm}L{1.0cm}L{2.0cm}L{1.55cm}@{}}
\toprule
\textbf{Work} & \textbf{Tool} & \textbf{Schedulers} & \textbf{Focus} \\
\midrule
Prosa~\cite{prosa} & Rocq & EDF, fixed-priority & schedulability \\
\addlinespace[2pt]
FIFO RTA~\cite{fiforta} & Coq & FIFO (real-time) & response time \\
\addlinespace[2pt]
seL4~\cite{sel4} & Isabelle & kernel RR & kernel correctness \\
\addlinespace[2pt]
CertiKOS~\cite{certikos} & Coq & generic scheduler & kernel correctness \\
\addlinespace[2pt]
RT-CertiKOS~\cite{rtcertikos} & Coq & fixed-priority & real-time scheduling \\
\addlinespace[2pt]
MLFQ~\cite{mlfq-tla} & TLA+ & MLFQ & formal spec. \& vis. \\
\addlinespace[2pt]
\textbf{This work} & Rocq & FIFO, SJF, Lottery, MLFQ, Stride, CFS & policy verification \& reusable patterns \\
\bottomrule
\end{tabular}
\end{table}


\section{Modeling Approach} \label{SEC:Modeling_frame}

This section describes the general structure of the Rocq models developed in this work. The models capture scheduling policies at the algorithmic level rather than complete operating-system implementations. Consequently, they abstract away implementation-specific aspects such as interrupts, device I/O, memory management, synchronization, and multicore execution, while retaining the information required to model scheduler behavior and reason about its correctness.

Each scheduler model consists of three main components:

\begin{itemize}
\item \textbf{Process representation}: the information maintained for each process, including attributes such as arrival time, estimated burst length, priority, ticket count, pass value, or virtual runtime.

\item \textbf{Scheduler state}: the data structures used by the scheduler to make scheduling decisions, such as waiting lists, priority queues, runnable sets, or runqueues.

\item \textbf{Scheduling step}: the transition rule that selects the next process for execution and, where applicable, updates the scheduler state.
\end{itemize}

Although all scheduler models follow this common organization, their scheduling steps differ according to the semantics of the scheduling policy. For FIFO, SJF, Stride, MLFQ, and the list-based CFS model, scheduling is deterministic and is represented by a total function mapping one scheduler state to the next:

\[
\texttt{tick} : \mathit{sched\_state} \to \mathit{sched\_state}
\]

Lottery scheduling differs because process selection depends on a ticket draw. Consequently, multiple valid scheduling outcomes may exist for the same scheduler state. Rather than modeling scheduling as a deterministic function, Lottery is represented as a relation between a scheduler state and a selected process:

\[
\texttt{tick} : \mathit{lottery\_state} \to \mathit{proc} \to \mathit{Prop}
\]

The red-black-tree implementation of CFS follows a third modeling style. Since the runqueue may be empty, the scheduling step returns an optional result. When a runnable process exists, the model returns both the selected process and the updated scheduler state:

\[
\texttt{tick} : \mathit{cfs\_rbt\_state} \to
\mathit{option}~(\mathit{proc} \times \mathit{cfs\_rbt\_state})
\]

Table~\ref{tab:model_state} summarizes the principal state components maintained by each scheduler model together with the corresponding process-selection criterion. The detailed formalization of each scheduler is presented in Section~\ref{SEC:Schedulers}.

\begin{table}[!ht]
\caption{Main State Information Used by Each Scheduler Model}
\label{tab:model_state}
\centering
\footnotesize
\begin{tabular}{@{}lll@{}}
\toprule
\textbf{Scheduler} & \textbf{Main state} & \textbf{Selection basis} \\
\midrule
FIFO & waiting and finished lists & arrival time \\
SJF & waiting and finished lists & burst length \\
Lottery & runnable list & ticket interval \\
MLFQ & priority queues and boost timer & highest non-empty queue \\
Stride & runnable list & pass value \\
CFS list & runnable list and minimum runtime & virtual runtime \\
CFS RB & red-black-tree runqueue & tree minimum \\
\bottomrule
\end{tabular}
\end{table}

These modeling choices explain why the subsequent correctness proofs are not identical across all schedulers. While several properties can be formulated uniformly, others depend on the underlying representation of the scheduling step. For example, determinism follows directly for function-based schedulers, whereas Lottery requires separate proofs that characterize the possible outcomes of different valid ticket draws. Despite these differences, recurring verification patterns emerge across scheduler families. These patterns capture common proof structures that can be reused across multiple scheduling policies and are introduced later in the paper.

\section{Formal Scheduler Models} \label{SEC:Schedulers}

The previous section introduced the common modeling framework used throughout this work. This section presents the scheduler-specific formalizations, emphasizing the key modeling ideas, machine-checked results, and verification patterns that distinguish the scheduler families. Owing to space limitations, only the principal modeling ideas are presented here. The complete Rocq formalizations and machine-checked proofs are available in the artifact\footnote{\url{https://codeberg.org/cascade/sched-fv}}.


\subsection{Ordering-Based Schedulers}

FIFO, SJF, Stride, and the list-based CFS model form the ordering-based scheduler family. Although these schedulers optimize different scheduling objectives, they all select the runnable process with the smallest value of a scheduler-specific ordering key: arrival time for FIFO, CPU burst length for SJF, pass value for Stride, and virtual runtime for CFS. Consequently, their correctness arguments follow the same minimum-key verification pattern.

In addition to establishing correct process selection, we prove structural invariants showing that scheduler transitions preserve the set of process identifiers. Thus, scheduling steps neither create nor remove processes while updating the scheduler state. These results form the basis of the reusable minimum-key verification pattern presented later.


\subsection{Multi-Level Feedback Queue (MLFQ)}

\begin{figure}[!ht]
\centering
\begin{tikzpicture}[>=Latex,
  every node/.style={font=\footnotesize},
  q/.style={draw, minimum width=4.2cm, minimum height=8.5mm},
  t/.style={draw, rounded corners=0.4pt, minimum width=5mm, minimum height=5mm,
            inner sep=0pt, font=\scriptsize, fill=black!5}]
\node[q] (q2) at (0,1.9) {};
\node[q] (q1) at (0,0.95) {};
\node[q] (q0) at (0,0.0) {};
\node[anchor=west, font=\scriptsize] at ([xshift=1.5mm]q2.west) {Q2 (high)};
\node[anchor=west, font=\scriptsize] at ([xshift=1.5mm]q1.west) {Q1 (mid)};
\node[anchor=west, font=\scriptsize] at ([xshift=1.5mm]q0.west) {Q0 (low)};
\node[t] at ([xshift=-16mm]q2.east) {P1};
\node[t] at ([xshift=-9.5mm]q2.east) {P4};
\node[t] at ([xshift=-9.5mm]q1.east) {P2};
\node[t] at ([xshift=-16mm]q0.east) {P3};
\node[t] at ([xshift=-9.5mm]q0.east) {P5};
\draw[->] (q2.east) to[out=0,in=0]
  node[right=1pt,font=\scriptsize]{demote} (q1.east);
\draw[->] (q1.east) to[out=0,in=0] (q0.east);
\draw[->,thick] (q1.west) to[out=180,in=180,looseness=1.1] (q2.west);
\draw[->,thick] (q0.west) to[out=180,in=180,looseness=1.7] (q2.west);
\node[anchor=east, align=center, font=\scriptsize] at ([xshift=-15mm]q1.west)
  {boost\\(every period)};
\end{tikzpicture}
\caption{Illustration of the MLFQ Scheduling Strategy}
\label{fig:mlfq}
\end{figure}
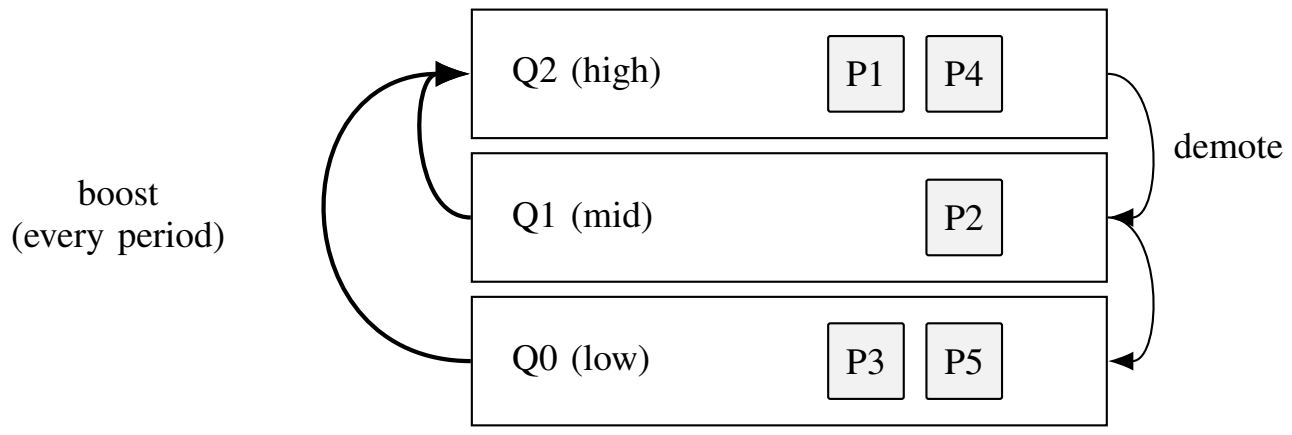

Unlike the ordering-based schedulers, MLFQ does not rely on a single ordering key. Instead, runnable processes are organized into multiple priority queues, and scheduling decisions are made from the highest non-empty queue. Figure~\ref{fig:mlfq} illustrates this organization. During a scheduling step, the model either performs a global priority boost or schedules the process at the head of the highest-priority non-empty queue. After execution, the process is either reinserted into the same queue or demoted after exhausting its allotted time slice.

We prove that a priority boost returns every runnable process to the highest-priority queue, establishing the correctness of the boost mechanism. Unlike the ordering-based schedulers, the proof uses reasoning about queue hierarchy rather than key ordering, motivating the layered-priority verification pattern. A complete starvation-freedom proof remains future work.


\subsection{Lottery Scheduling}

Lottery scheduling allocates processor time using tickets rather than an ordering key. Each runnable process owns a number of tickets, and a valid draw is any natural number below the total ticket count. The selected process is the one whose ticket interval contains the draw. Since different valid draws may select different processes from the same scheduler state, Lottery is modeled as a relation rather than a deterministic function.

We prove that every draw within a process's ticket interval selects that process and that each ticket interval contains exactly as many draw values as the corresponding process owns tickets. Together, these results establish the formal relationship between ticket ownership and process selection, forming the ticket-interval verification pattern.


\subsection{CFS with a Red-Black-Tree Runqueue}

The list-based CFS model belongs to the ordering-based scheduler family because it follows the minimum-key verification pattern based on virtual runtime. We additionally formalize a CFS variant whose runqueue is implemented using Rocq's verified finite-set library based on red-black trees~\cite{msets}, demonstrating that the verification framework is independent of the underlying data representation.

We prove that each scheduling step selects the minimum virtual-runtime process while preserving the runqueue invariant. When the runqueue is non-empty, the verified scheduler returns both the selected process and updated scheduler state.

\section{Comparative Verification of Scheduler Properties}
\label{SEC:properties}

This section summarizes the verified properties established for the scheduler models. Tables~\ref{tab:headline} and~\ref{tab:struct} classify the verified results. A check mark (or ``yes'') denotes a proved property, ``no'' denotes a property that does not hold, ``partial'' denotes a verified fragment of a broader property, ``open'' indicates that the property has not yet been proved, and ``n.a.'' denotes that the property is not applicable to the corresponding scheduler model. All reported results are mechanically verified by Rocq without admitted lemmas or additional axioms.

\subsection{Headline Properties}

Table~\ref{tab:headline} summarizes the scheduler-level properties. Starvation freedom concerns whether waiting jobs eventually execute, determinism whether identical scheduler states always produce the same outcome, fairness whether scheduling follows the intended resource-allocation policy, and work conservation whether runnable work always permits scheduling progress.

The headline results should be interpreted as follows. ``Batch'' for FIFO and SJF assumes that all jobs are initially present in the waiting list. Lottery fairness refers to the verified ticket-interval property, while work conservation assumes that all runnable processes hold positive ticket counts. For MLFQ, ``partial'' reflects the verified priority-boost mechanism but not a complete starvation-freedom or work-conservation argument. The two CFS rows distinguish the list-based and red-black-tree implementations.

\vspace*{-0.4cm}

\begin{table}[H]
\caption{Verified Headline Scheduler Properties}
\label{tab:headline}
\centering
\footnotesize
\begin{tabular}{@{}L{1.0cm}L{1.35cm}L{1.35cm}L{1.25cm}L{1.65cm}@{}}
\toprule
\textbf{Sched.} &
\textbf{Starvation freedom} &
\textbf{Determinism} &
\textbf{Fairness} &
\textbf{Work conservation} \\
\midrule
FIFO & yes (batch) & yes & n.a. & yes \\
SJF & yes (batch) & yes & open & yes \\
Lottery & open & no & partial (tickets) & yes (positive tickets) \\
MLFQ & partial (boost) & yes & open & partial (non-boost) \\
Stride & open & yes & open & yes \\
CFS list & open & yes & open & yes \\
CFS r-b & open & yes & open & yes \\
\bottomrule

\end{tabular}

\end{table}

\vspace*{-0.2cm}

Figure~\ref{lst:headline_example} illustrates a representative scheduler-level theorem. It states that, in the FIFO batch model, executing the scheduler for the length of the waiting list empties that list.

\vspace*{-0.2cm}

\begin{figure}[H]
\begin{lstlisting}
Theorem fifo_starvation_free : forall s,
    (run (length s.(waiting)) s).(waiting) = [].
\end{lstlisting}
\caption{FIFO Starvation Freedom in the Batch Model}
\label{lst:headline_example}
\end{figure}

\vspace*{-0.3cm}

\subsection{Structural Properties}

Table~\ref{tab:struct} reports one-step structural properties. Selection establishes that a scheduler chooses the process prescribed by its policy. Conservation and uniqueness concern preservation of process identifiers. Monotonicity states that the scheduler's progress measure does not decrease, while invariant preservation ensures that scheduler-specific validity conditions are maintained across scheduling steps.

\vspace*{-0.2cm}

\begin{table}[H]
\caption{Verified Structural Scheduler Properties}
\label{tab:struct}
\centering
\footnotesize
\begin{tabular}{@{}L{1.0cm}ccccc@{}}
\toprule
\textbf{Sched.} & \textbf{Select} & \textbf{Conserv.} &
\textbf{Uniq.} & \textbf{Monot.} & \textbf{Invar.} \\
\midrule
FIFO & \yes & \yes & \yes & \yes & \na \\
SJF & \yes & \yes & \yes & \yes & \na \\
Lottery & \yes & \na & \na & \na & \na \\
MLFQ & \yes & \yes & \yes & partial & \na \\
Stride & \yes & \yes & \yes & \yes & \na \\
CFS list & \yes & \yes & \yes & \yes & \yes \\
CFS r-b & \yes & open & open$^{a}$ & partial & \yes \\
\bottomrule
\end{tabular}

\vspace{2pt}
\footnotesize{$^{a}$The red-black-tree ordering uses virtual runtime, process identifier, and weight. The current tree-backed model proves selection and red-black invariants, but does not include a separate global pid-uniqueness theorem.}
\end{table}

The structural results require only a few additional remarks. Lottery reports ``n.a.'' for state-based properties because its scheduling step returns only the selected process. FIFO, SJF, MLFQ, and Stride do not maintain an explicit state-validity predicate, so invariant preservation is not applicable. For MLFQ, monotonicity is partial because only the boost-timer behavior is currently verified. The CFS list model preserves its explicit state-validity conditions, whereas the current red-black-tree model establishes runqueue invariants but leaves separate process-identifier preservation theorems as future work.

Figure~\ref{lst:structural_example} presents a representative structural theorem. It establishes that a successful scheduling step selects the minimum virtual-runtime process while preserving the red-black-tree runqueue invariant.

\vspace*{-0.2cm}

\begin{figure}[H]
\begin{lstlisting}
Theorem tick_run_selects_min_and_preserves_rbt :
  forall s selected next_state,
    tick s = Some (selected, next_state) ->
    Runqueue.min_elt s.(runqueue) = Some selected /\
    Runqueue.In (update_proc selected)
        next_state.(runqueue) /\
    rbt_invariants next_state.(runqueue).
\end{lstlisting}
\caption{CFS Red-black-tree Selection and Invariant Preservation}
\label{lst:structural_example}
\end{figure}

\section{Shared Proof Patterns} \label{SEC:Interfaces}

The previous sections introduced the scheduler models and their verified properties. We now compare these results to identify recurring proof patterns shared across different scheduler families. Although the schedulers implement different scheduling policies, many of their correctness arguments follow a small number of common proof structures.

Table~\ref{tab:interfaces} summarizes the proof patterns identified in this work. Rather than forcing all schedulers into a common implementation model, these patterns capture the common correctness arguments while allowing each scheduler to retain its own state representation and scheduling semantics.

\vspace*{-0.3cm}

\begin{table}[H]
\caption{Proof Patterns Identified Across the Scheduler Models}
\label{tab:interfaces}
\centering
\scriptsize
\begin{tabular}{@{}cL{1.35cm}L{1.95cm}L{2.15cm}@{}}
\toprule
\textbf{No.} & \textbf{Pattern} & \textbf{Schedulers} & \textbf{Checked claim} \\
\midrule
1 & Selection safety & Lottery, CFS RB & selected process is runnable \\
2 & Minimum key & FIFO, SJF, Stride, CFS list & selected key is minimal \\
3 & Layered priority & MLFQ & selected layer is highest available \\
4 & Ticket interval & Lottery & draw selects ticket owner \\
\bottomrule
\end{tabular}
\end{table}

\vspace*{-0.2cm}

Selection safety forms the foundation of the remaining proof patterns by establishing that every selected process is runnable. The minimum-key pattern applies to FIFO, SJF, Stride, and the list-based CFS model, where correctness is established by proving that the selected process minimizes a scheduler-specific ordering key. MLFQ follows the layered-priority pattern, requiring proof that scheduling always occurs from the highest non-empty priority queue. Lottery follows the ticket-interval pattern, which establishes that every valid draw selects the owner of the corresponding ticket interval. Together, these proof patterns demonstrate that diverse scheduling policies can be verified using a small number of reusable correctness arguments.

Figure~\ref{lst:shared_patterns} shows representative Rocq predicates that formalize the proof patterns summarized in Table~\ref{tab:interfaces}. Individual scheduler models instantiate these predicates to establish their corresponding correctness results.

\vspace*{-0.3cm}

\begin{figure}[H]
\begin{lstlisting}
Definition selects_from_runnable S : Prop := forall s p,
    S.(sel_select) s p -> S.(sel_runnable) s p.

Definition min_key_correct S : Prop :=
  min_key_selects_from_runnable S /\ selects_min_key S.

Definition layered_correct S : Prop :=
  layered_selects_from_runnable S /\ selects_highest_layer S.

Definition ticket_interval_correct S : Prop :=
  owned_draws_select_owner S /\
  owned_draw_count_matches_share S.
\end{lstlisting}
\caption{Representative Rocq Predicates for the Shared Proof Patterns}
\label{lst:shared_patterns}
\end{figure}


\section{Limitations and Future Work} \label{SEC:limits}

This work focuses on scheduler policies rather than complete operating-system implementations. The models capture scheduler state, runnable processes, and one scheduling step, while abstracting away interrupts, device I/O, memory management, locking, preemption, and multicore execution. This level of abstraction keeps the models sufficiently compact to formalize and compare multiple scheduling policies within a single Rocq development. Consequently, the reported results should be interpreted as policy-level correctness guarantees rather than kernel-level verification results.

Some verification results also reflect the assumptions of the underlying models. FIFO and SJF establish starvation freedom only for the batch-processing setting in which all runnable processes are initially available. Lottery scheduling verifies the correctness of individual ticket draws rather than long-run probabilistic fairness. Similarly, the current MLFQ formalization establishes the correctness of periodic priority boosting without deriving starvation-freedom guarantees for arbitrary executions. Finally, the Stride and CFS models verify local scheduling and accounting properties, whereas complete proportional-share guarantees and a formal refinement between the list-based and red-black-tree CFS models remain outside the scope of the current development.

These limitations naturally suggest several directions for future work. The most immediate extensions include establishing complete starvation-freedom guarantees for MLFQ, incorporating probabilistic reasoning for repeated Lottery draws, proving long-run proportional-share properties for Stride and CFS, and developing a refinement proof relating the two CFS models. More broadly, the proposed proof patterns could be applied to additional scheduler designs, including hybrid schedulers that combine layered priorities with minimum-key selection, thereby extending applicability of the framework.


\section{Conclusion} \label{SEC:conclusions}

This paper presented a comparative Rocq formalization of six representative general-purpose CPU scheduling policies: FIFO, SJF, Lottery, MLFQ, Stride, and CFS. In addition to a list-based CFS model, we formalized a variant whose runqueue is implemented using Rocq's verified red-black-tree library. For each scheduler, we established machine-checked correctness results and classified the verified properties as proved, partial, open, or not applicable.

Beyond verifying the individual schedulers, we identified recurring proof patterns that characterize their correctness arguments. FIFO, SJF, Stride, and the list-based CFS model share a minimum-key proof pattern, MLFQ follows a layered-priority proof pattern, and Lottery follows a ticket-interval proof pattern. These proof patterns provide a common basis for comparing scheduler families while allowing each scheduler to retain its own mathematical model and operational semantics. We hope that this framework will facilitate the formal verification of additional scheduling policies in future work.

\bibliographystyle{IEEEtran}
\bibliography{references}

\end{document}